\documentstyle[prd,aps,preprint]{revtex}
\tightenlines
\newcommand {\be}{\begin{eqnarray}}
\newcommand {\ee}{\end{eqnarray}}
\input epsf.tex
\def\DESepsf(#1 width #2){\epsfxsize=#2 \epsfbox{#1}}
\begin{document}

\draft
\preprint{\vbox{
\hbox{UMD-PP-99-085}
\hbox{SMU-HEP-99-01}
}}
\title{Mirror Matter MACHOs}
\author{ Rabindra N. Mohapatra$^1$ and Vigdor L. Teplitz$^2$ }

\address{$^1$Department of
Physics, University of Maryland, College Park, MD, 20742\\
$^2$ Department of Physics, Southern Methodist University, Dallas, TX-75275.}
\date{February, 1999}
\maketitle
\begin{abstract}

We propose that the massive compact halo objects (MACHOs) observed in the 
recent microlensing experiments with an apparent best fit mass of about 
$0.5 M_{\odot}$ are objects made out of ``mirror'' baryonic matter rather 
than familiar baryons. Such a possibility arises naturally 
within the framework of mirror matter models proposed recently to 
accomodate the sterile neutrinos that seem necessary to solve 
all the neutrino puzzles simultaneusly. We show that for mirror 
matter parameters that fit the neutrino observations, the maximum mass of 
mirror stars are of order $0.5 M_{\odot}$ and their main sequence 
lifetime is much less than the age of the universe. They are therefore 
likely to be black holes. Mirror matter machos have the advantage 
that they do not suffer from the problems 
encountered in the conventional red, brown or white dwarf interpretation. 
The calculations also apply to the question of how the world  of familiar 
matter would be different if all fundamental mass parameters were.
\end{abstract}

\section{Introduction}

The nature of the dark matter in the universe, for which there is 
considerable observational evidence, is a mystery\cite{book}.
There are a number of experiments in progress to resolve this mystery.
Here we will address the issues raised by the 
microlensing experiments\cite{micro} which monitor millions of stars in the 
neighbouring Large Magellanic Cloud (LMC) to see if any them suddenly
brighten for a certain duration of time and then fade away again.
The brightening observed is attributed to a lensing effect due to 
passage of a dark massive object, presumably from our galaxy, in front of 
the LMC star. The duration $\Delta t$ of the brightening of a MACHO event
has been calculated and is proportional to 
$\sqrt{m}/v$ where $m$ is the mass of the object and $v$ is its velocity.
If one assumes that the object is from our galaxy, its velocity is 
determined; as a result, from $\Delta t$, one can 
deduce the macho mass. Based on the 14 events from the MACHO and the 
EROS collaborations that are attributable to MACHO events, one obtains a 
best fit mass\cite{micro} of $0.5M_{\odot}$ for the MACHOs. Also it has been
established that there are no MACHO candidates with masses between 
$10^{-7}M_{\odot}$ to $10^{-2}M_{\odot}$. It is also expected that these 
objects can comprise as much as 30 to 50\% of the halo mass 
fraction\cite{gyuk}. The question then arises: ``what are these objects?''.

The simplest possibilities are conventional baryonic objects
such as red, brown or white dwarfs whose masses are expected to be in 
this range or that they are neutron stars. It has however been argued by 
Hegyi and Olive\cite{olive} that a large class of baryonic candidates are 
incompatible with observations. More recently, Freese et al\cite{freese}
have made a detailed study of the possibility that they could be red, brown 
or white dwarfs and have found such interpretations to be highly problematic.
(For a more recent analysis of some of these issues, see \cite{chab}.) We 
will summarize these difficulties in a subsequent section. Accepting 
them for the moment their conclusion then leads us to search for 
alternative explanations. Our finding in this paper is that if there 
exists a mirror universe with identical particle and force content to 
the visible universe prior to gauge symmetry 
breaking\cite{lee,khlopov,berezhiani}, then for certain 
choice of the symmetry parameters, the maximum mass of the ``mirror'' stars
is of order $0.5 M_{\odot}$ and they could therefore be the machos observed
in the microlensing experiments. Since they are made of mirror baryons, they
avoid all the problems encountered by machos made of conventional baryonic
matter\cite{blin,berezhiani}. We also estimate the main sequence lifetime of 
the mirror stars 
and find that in the parameter range of interest it is much less than the 
age of the universe. As a result, the machos are in the form of black holes.
Let us note that precisely such models\cite{berezhiani} have recently 
been proposed to accomodate all the neutrino observations by identifying
the lightest of the mirror neutrinos with the sterile neutrino needed in
understanding the LSND result together with the solar and the atmospheric 
neutrino results.

While this is the main result of our paper, our study also applies to the 
question of how the world of familiar matter would look if the masses are all
scaled by a common factor. Similar questions have been studied in the 
past\cite{barr,berezhiani,teplitz}.

The main results of this letter are the following: we review the scaling 
laws for maximum and minimum values for the stellar masses as the
masses of the electrons, W-bosons, protons and neutrons vary together 
($m_i\rightarrow \zeta m_i$). The maximum 
value, which is of particular interest for us is derived by using an argument
in the literature that the radiation pressure inside a stable compact 
stellar object should not exceed the gas pressure inside it and by finding 
how that condition scales as the elementary particle masses vary. We 
find that, for $\zeta$ of
order 15, mirror stars have maximum mass of the order of the $0.5 M_{\odot}$ 
needed
for MACHOs in the halo. We give qualitative arguments to support 
the hypothesis that the initial stellar mass function (IMF) for the mirror 
sector is likely to peak near the maximum value for $\zeta$ in the parameter 
region of interest to neutrino physics. These two arguments ($\zeta\sim 
15$ explains both MACHO masses and the neutrino results) in 
our opinion strengthen the conjecture that mirror matter machos are 
the dark massive halo objects in our galaxy seen in the microlensing data.

\section{Scaling laws for stellar masses and main sequence lifetimes}

We begin with a brief overview of the mirror universe model and the 
the parameters describing fundamental forces in the mirror sector. As 
mentioned, one considers a duplicate version of the standard model
with an exact mirror symmetry which is broken in the process of gauge 
symmetry breaking.
All particles and parameters of the mirror sector will be denoted by
a prime over the corresponding familiar sector symbol- e.g. mirror quarks
are $u',d',s',$ etc and mirror Higgs field as $H'$, mirror QCD scale as 
$\Lambda'$ . We assume that $<H'>/<H>=\Lambda'/\Lambda\equiv
\zeta$. Since one expects the masses of the neutron and proton to be 
given by the scale $\Lambda$ and charged lepton (and current quark) masses
to be given by $<H>$, scaling both parameters by the same amount implies that
all fermion masses that are relevant to the discussion of stellar structure
scale by the same amount, i.e. we have
$m_i\rightarrow\zeta\,m_i$ with $i=n,p,e,W,Z.$  Furthermore, this gives weak 
cross sections varying as
$\zeta^{-4}$ for fixed values of energy.  With these simple rules, assuming that
electroweak and strong coupling constants do not change, we can say a great
deal about how the properties of stars would change.

We start with the four equations of stellar structure:

\be\label{st1}
dP/dr=-G\rho(r)M(r)/r^2
\ee
\be\label{st2}
dM(r)/dr=4\pi\,r^2\rho(r)
\ee
\be\label{st3}
L(r)/4\pi\,r^2=-(16/3)\sigma_{SB}(T^3/\rho\kappa)dT/dr
\ee
\be\label{st4}
dL/dr=4\pi\,r^2\epsilon(r)\rho(r)
\ee
where $\kappa(r)$ is the opacity (cross section per unit mass) at radius $r$,
$\sigma_{SB}$ the Stefan-Boltzmann constant, L(r) the luminosity at 
radius $r$,
and $\epsilon(r)$, the rate of energy generation per unit mass at radius 
$r$. 
We will need three terms in the equation of state (below) taken one or two
 at a time:
\be\label{eos}
P=(\rho/m)kT + (4\sigma_{SB}/3c)T^4 + (h^2/2m_e)(3/8\pi)^2(\rho/m)^{5/3}
\ee

where the three terms represent gas pressure, radiation pressure, and
(non-relativistic) degenerate electron pressure.  $m$ is the nucleon mass, $m_e$
that of the electron.  We have neglected such niceties
as keeping track of how many objects there are for each $m$ of gas (2 for
 H,
3/4 for He, etc)

We will make standard, illuminating if crude, approximations \cite{clayton}
 in order to understand the $\zeta$ behavior of the
solutions to the above equations.  First we write

\be\label{peq}
P=\rho\,GM/R,\ \ \ \rho=3M/4\pi\,R^3
\ee
where $P$ and $\rho$ are roughly core averages. Here $M$ and $R$ are mass
 and radius of core or star; our approximations are not good enough to be
 precise on such points.(In practice we will adjust $M$ to be the mass of the
whole star and $R$ will fall short of the core radius for the sun).  
Equation (\ref{peq})
gives the useful relation

\be\label{prho}
P=(4\pi/3)^{1/3}GM^{2/3}\rho^{4/3}
\ee

To find the minimum mass of a star, we neglect the radiation pressure term in
Equation (\ref{eos}), insert into Equation (\ref{prho}), solve for T and
maximize with respect to $\rho$, giving

\be\label{tmax}
kT = (G^2/2)(8\pi/3)^{2/3}M^{4/3}m^{8/3}m_e/h^2
\ee
Following Phillips \cite{clayton}, we set $T=T_{ig}$, the lowest 
temperature 
that gives sufficient burning to match energy escape and solve for $M$, 
obtaining

\be\label{mmin}
M_{min}\sim[h^2kT_{ig}/(m_eG^2m^{8/3})]^{3/4}
\ee
We know from more detailed analysis that $M_{min}$ is of the order of
$0.07M_{\odot}$ and $T_{ig}\sim10^6K$.  We use Equation (\ref{mmin}) to 
obtain the variation with $\zeta$. $m$ and $m_e$ go as $\zeta$.  $T_{ig}$,
 in
principle must be found by solving the four coupled Equations
(\ref{st1}-\ref{st4}).  Roughly, however, nuclear binding energies will go
linearly with $\zeta$ so we will approximate the variation of $T_{ig}$ as
linear as well.  We will see below that the solution of approximate 
equations
gives $T$ varying with $\zeta$ (numerically) not greatly different.  We thus
obtain

\be\label{mminvary}
M_{min}\sim\zeta^{-2}
\ee

We can also, again following Phillips \cite{clayton}, use Equation 
(\ref{eos}) to find
the maximum mass of a (main sequence) star.  As  the mass of the star 
gets bigger, the core temperature rises. Therefore, of the three 
terms in the expression for the pressure in the Equation \ref{eos}, 
we expect $P_g$ and $P_r$ to dominate. Following Phillips\cite{clayton},
we parameterize them as fractions of the total pressure $P$ as below: 

\be\label{beta}
P_g=\beta\,P,\ \ \ P_r=(1-\beta)P
\ee
We eliminate $T$ and solve Equation (\ref{beta}) for P, obtaining

\be\label{pbet}
\beta P=[(\rho\,k/m)^4(\beta^{-1}-1)/(4\sigma_{SB}/3c)]^{1/3}
\ee
Using Equation (\ref{prho}) again then gives
\be\label{mmax}
M_{max}\sim[(1-\beta)c/\sigma_{SB}]^{1/2}G^{-3/2}(k/m)^2/\beta^2
\ee
As $\beta$ approaches $1$, the energy density is increasingly dominated by
photons (relativistic particles) and stars become unstable.  Taking a 
cutoff around $\beta\sim1/2$ gives a maximum stellar mass around 
$70M_{\odot}$. Thus
the range for stars is roughly $0.07M_{\odot}$ to $70M_{\odot}$.  From 
Equation
(\ref{mmax}) one sees, in the approximation that instability sets in at the
same $\beta$ independent of $\zeta$, that $M_{max}$ varies as $\zeta^{-2}$
(like $M_{min}$).  It is similarly easy to see from the standard 
expression for
the Chandrashekar mass that it too varies as $\zeta^{-2}$.  Note that, in a
model with $m_e$ varying linearly with $\zeta$, but $m$ constant, both 
$M_{CH}$
and $M_{max}$ would be independent of $\zeta$ while $M_{min}$ would go as
$\zeta^{-3/4}$ (since higher mass for the electron would permit contraction to
higher densities before Pauli repulsion becomes important).  Note that 
such is
the case for the mirror matter model investigated in references 
\cite{berezhiani} in order to solve the neutrino puzzles.

We now consider stellar burning as $\zeta$ varies.  We approximate Equation
(\ref{st3}) as
\be\label{lum}
L=(16\pi/3)^2\sigma_{SB}(RT)^4/(\kappa\,M)
\ee
where $\kappa$ is the opacity, for which we keep just $\gamma-e$ scattering
contribution i.e. we take $\kappa$ as\footnote{Omitting other 
contributions to opacity tends to overstimate $\kappa$ as $\zeta$ 
increases and hence underestimate the luminosity and overestimate the 
main sequence lifetime. For our purpose, therefore, our assumption about
$\kappa$ is a conservative one.}

\be\label{opac}
\kappa=\sigma_T/m_e\sim\zeta^{-3}\kappa_{\odot}
\ee

Since the rate of energy generation is determined by the weak interaction
 rate
for $p+p\rightarrow\,e^++d+\bar{\nu}$, we approximate Equation(\ref{st4}) by

\be\label{lumap}
L=\epsilon\,M
\ee
with

\be\label{eps}
\epsilon=\overline{\sigma\,v}E_{pp}\rho/m^2
\ee

We take $\overline{\sigma\,v}$ from Clayton \cite{clayton} and, as 
above, continue in
$\zeta$, obtaining

\be\label{eps2}
\epsilon=(\rho\,E_{pp}/m^2)\zeta^{-3}f(T_6/\zeta)
\ee
where $T_6=T/10^6$, $E_{pp}$ is the energy release for the full $pp$ 
chain into
other than neutrinos, and

\be\label{ftil}
f(x)=3\times10^{-37}x^{-2/3}e^{-33.81x^{1/3}}[1+0.021x^{-1/3}+0.01x^{2/
3}+9.5\,x\times10^{-4}]
\ee

One factor of $\zeta^{-1}$ in Equation (\ref{eps2}) comes from the 
parenthesis preceding $\zeta^{-3}$ and two come from the behavior of the 
weak 
cross section, taking into account energies increasing with $\zeta$.  

Equating the two expressions above for $L$ permits us to solve for $RT$ as a
function of $T$, while Equations (\ref{eos}) and (\ref{prho}) give
$RT=MGm/k$. Combining these results we can write,

\be\label{m4}
M^4=4(3/16\pi)^3(k/mG)^7(\epsilon_{\odot}\kappa_{\odot}/
\rho_{\odot}\sigma_{SB})T^3\zeta^{-13}f(T_6/\zeta)/f(15)
\ee
where we have normalized to the temperature ($T_6=15$) at the center of
 the
sun.

Equation (\ref{m4}) gives an approximation to the variation of stellar masses
with $\zeta$.  With it, we can solve for $R$, $\rho$, $L$, and the main
sequence lifetime $t_{MS}=(0.1Mc^2/L)$.  In Figure 1 (a,b) we give the 
results, as a function of $M$, for temperature $T$, radius $R$, and main
sequence lifetime, $t_{MS}$ for $\zeta=1.0, 15$.  We have
inserted an overall factor to scale the main sequence lifetime of the sun
($\zeta=1, <T_6>\sim7.5$) to $10^{10}$ years.  The scale units are $10^6K$,
$10^{10}cm$, and $10^9\,years$ for the three quantities.  It should be 
noted
that, in the approximations made, the solar (core) radius comes out to about
$0.3\times10^{10}cm$ while, in real life, two thirds of the sun's mass 
(with
temperature $T_6>7$) extends out to about $1.5\times10^{10}cm$.

We see from Figure (1) that the ranges of radii and main sequence lifetimes
fall with increasing $\zeta$ while that of temperatures increases.  In Figure
(2) we address this increase in more detail by plotting $T$ against 
$\zeta$ for
four values of an index that runs from 1-100 as $M$ varies from $M_{min}$
 to
$M_{Max}$ in equal logarithmic increments.  Roughly, we see that $T$ goes as
$\zeta^{4/3}$, so that the assumption above that $T_{ig}\sim\zeta$ is not
grossly out of line.  It should be noted that, for massive stars (with
$\zeta=1$), the pp cycle on which the above considerations are based is
replaced by CNO cycles (Clayton, ref.\cite{clayton}), in which C, N, 
and O catalyze He
production from H at high enough temperatures.  Since the weak interactions in
the cycle are all weak decays, which increase, in rate, linearly with 
$\zeta$, we expect that massive stars, for large $\zeta$, should have even
shorter main sequence lifetimes than those of Figure 1.  Thus the lifetimes in
Figure 1 should be considered upper bounds.  However estimating the abundances
of C, N, O could be quite complicated (see below).

We will use, below, the unsurprising result from Figure (1), that for 
$\zeta>5$
main sequence lifetimes are all much shorter than the age of the universe;
 they
fall roughly as $\zeta^{-3}$.

\section{Mirror vs Baryonic MACHOs}

In this section we show that mirror matter MACHOs (MMMs) provide a good 
explanation of
microlensing events within which microlensing data can determine the parameter 
$\zeta$.  Finally, we discuss briefly tests of the model.

First, mirror matter resolves a number of MACHO problems.  Fields, Freese
 and Graff \cite{freese}, in a very detailed work, raise several 
problems with baryonic MACHO candidates, including:

--All baryonic candidates require that the MACHO population be near the
minimum of the range permitted by observations and that $\Omega_B$ be near the
maximum of the range permitted by BBNS theory if the sum of individual baryon
components is to be less than the total baryon number.  MMMs
avoid this problem completely since mirror matter does not enter into the
baryon budget.

--Neutron stars and black holes from supernovae cannot fit the
fact that lensing events point to MACHO masses around $0.5M_{\odot}$. 
We will see below that mirror matter does provide an explanation of the mass
observation.

--Brown dwarf explanations conflict with a growing number of
observations that show that the index $\beta$ in the initial stellar mass
function (IMF), $N(m)\sim\,m^{-\beta}$, which is over 2 (2.35 is a commonly
accepted value \cite{adams}) for $m>M_{\odot}$ whereas it is
well under 2 for $m<M_{\odot}$; under 2 means that most of the mass is in
 the higher mass stars.  The conflict is because the frequency of 
microlensing
events appears to require a MACHO population with a total mass on the order of
up to half $\Omega_B$ while the decrease in $\beta$ for low masses precludes
such a result.  Since mirror matter is not baryonic, it has no such problem.

--Finally, the ``favorite candidate," white dwarfs, suffer from several
problems raised by Fields, Freese and Graff \cite{freese}, including: (i) 
not being
seen -- for example in the Hubble deep field (see Flynn, Bahcall,
and Gould \cite{flynn}) as they should be since $0.5M_{\odot}$ 
dwarfs can
 only
cool slowly; (ii) a need for a large population of galactic massive stars and 
their
supernovae to produce galactic winds to cleanse the galaxy of the processed
material in the white dwarf ejecta; and (iii) a contradiction between the 
amount of
carbon the progenitors would produce and the amount observed.  None of these
problems would arise with MMM progenitors.

We turn now to the question of why the MMMs should have
masses around $0.5M_{\odot}$.  This would be the case (see Figure (1)) if
 we have $\zeta$ greater than say, 15, since the maximum stellar mass then 
falls in
 the region between 0.5 and 1.0 solar masses.  The work of 
\cite{berezhiani} shows in
some detail that such values are just what is required to provide simultaneous
solution of the atmospheric, solar, and LSND neutrino problems (and provide
warm dark matter in addition).  There are, furthermore, reasons why (a) 
the
mirror matter stellar IMF would peak near the maximum mass and (b) remnant
masses would be similar to initial masses.  Both stem from the decrease in
cross sections with increasing $\zeta$.

Current theories of star formation (see, for example, Adams and Fatuzzo
\cite{adams} and references therein) from molecular cloud core collapse 
require
a mechanism to stop accretion during collapse, i.e. to limit the size of 
the
star.  Such mechanisms are based on scattering, but cross sections for
scattering of molecules, atoms, ions, and electrons off photons, atoms and 
molecules
will fall as $\zeta^{-2}$.  Thus it is not
unreasonable to expect that the mirror matter IMF should become more and 
more
strongly peaked near $M_{Max}$ as $\zeta$ increases.  Additionally, we might
expect a modest increase in $\zeta^2\,M_{max}$  in the
Section 2 estimate since scattering processes that create instability as
$\beta\rightarrow0$ will become less effective leading to smaller $\beta$
values and, through Equation (\ref{mmax}), larger values of $M_{max}$.

Cross section decrease also predicts that there should be little mass
loss between the initial star and the remnant.  As $\zeta$ increases, the
radius of the star decreases and neutrino cross sections decrease.   Thus
neutrino confinement times decrease sharply and it is doubtful that (Type
 II)
supernova shocks can be formed, thereby creating the population of $M_{Max}$
black holes that are detected as MACHOs.  These two $\zeta>15$ features --
decreased mass loss and mirror star masses peaking around $0.5M_{\odot}$ --
result in fixing $\zeta$ in a range optimal from the point of view of 
fitting neutrino results in reference \cite{berezhiani}.

In sum, MMMs appear to have a number of positive features as the
explanation of microlensing events.

Finally, we turn to the question of observational tests of the MMM
hypothesis.  The following come to mind: 

--Absence of any optical  observations of lensing objects as lensing
events accumulate;

--Within the qualitative picture above, relatively strong peaking of
lens masses into a narrow range;

--Possible detection of the black holes by some new method;
unfortunately estimates by Heckler and Kolb \cite{heck} show that,
even with new instruments (the Sloan digital sky survey telescope), black
 holes
under $10M_{\odot}$ could saturate the halo mass without being detectable
 from the signal from interstellar material infall. 

--Possible future detection of black hole MACH binaries\cite{nakamura} 
that if they exist in sufficient number through the emission 
of gravitational waves in experiments such as LIGO, VIRGO, TAMA and GEO.

\section{Lucky to Be Alive}

Finally we note some of the implications for the familiar world from
the above results on varying $\zeta$.  As noted in Section I, there is a
growing literature on the changes that would obtain if standard model, and
other parameters, were different\cite{barr,berezhiani,teplitz}.  The present 
investigation adds to those results.

The two most important changes as $\zeta$ grows would appear to be the
absence of supernovae and the decrease in stellar lifetimes.  These imply
 that,
as $\zeta$ grows, there would be lower abundances of heavy elements in the 
interstellar medium
with which to form planets and carbon based life forms on them, as well 
as less time in orbit around main sequence stars during which it would be 
possible for 
the latter to occur.  Although rates for radiative processes 
would increase  linearly with 
$\zeta$, main sequence lifetimes would fall as $\zeta^{-3}$ as shown in Figure 
1. It is the factor of $p^3$ in phase space that, apparently results in
increasing rates and decreasing cross sections.

As $\zeta$ decreases, the combination of decreasing rates and
increasing cross sections would be likely to interfere with current models of
star formation cited above.  For example, collapse times ($[G\rho]^{-1/2}
$)
would increase like $\zeta^{-2}$ while cross sections for scattering that
 would
tend to disperse the collapsing cloud would increase like $\zeta^{-2}$.

In conclusion, we have studied the variation of stellar mass and 
lifetimes as the masses of the elementary particles $m_e,m_p, m_n, M_W$
all vary in the same way (given by the parameter $\zeta$). We conclude that
for a value of $\zeta\simeq 15$, the maximum mass of the mirror stars
is around half a solar mass; as a result, they could be viable candidates
for the MACHOs observed in the various microlensing 
searches\footnote{Note that, in the mirror matter model of 
references\cite{berezhiani,teplitz} the neutron is unstable for $\zeta$ in 
the range of interest, so stellar structures become either mirror white 
dwarfs or black holes. Our considerations with respect to the maximum 
mass of such objects would still obtain. The result would be prompt black 
hole formation and the application to the MACHO problem would be the same 
as in this letter.}. The 
many problems encountered in trying to explain the $0.5 M_{\odot}$ machos 
white dwarfs and brown dwarfs etc are now easily avoided. We do
a crude analysis of the dependence of the ``main sequence'' life time of 
the  mirror stars as a function of the $\zeta$ variable and
find that for similar $\zeta$ values, the mirror macho is most likely a 
black hole since main sequence stellar lifetimes are few times shorter than 
the age of the universe. We also propose several tests of this hypothesis.

We are grateful to D. Clayton, S. Nussinov and D. Rosenbaum for 
many useful discussions. We are grateful to K. Freese and W. K. Rose for 
reading the manuscript and comments. The 
work of R. N. M. is supported by the National Science Foundation grant 
under no. PHY-9802551 and the work of V. L. T. is supported by the DOE  
under grant no. DE-FG03-95ER40908.

\vskip1.0in

\noindent{\bf Figure Caption}

\noindent Figure 1 (a, b): Temperature T, radius R
and the main sequence lifetime $t_{MS}$ of the mirror stars as a function 
of the stellar mass M for six different values of $\zeta=1.0, 15$.
 The units for the above are $10^6$K for T, $10^{10}$ cm for R and
$10^9$ yrs. for $t_{MS}$.

\noindent Figure 2. Variation of temperature T as a function of $\zeta$ for
4 of 100 equal mass steps between $M_{min}$ (step 1) and $M_{max}$ (step 
100). Step 3 for $\zeta=1$ corresponds to the sun.

\begin{figure}[htb] \begin{center}
\epsfxsize=9.5cm
\epsfysize=9.5cm
\mbox{\hskip -1.0in}\epsfbox{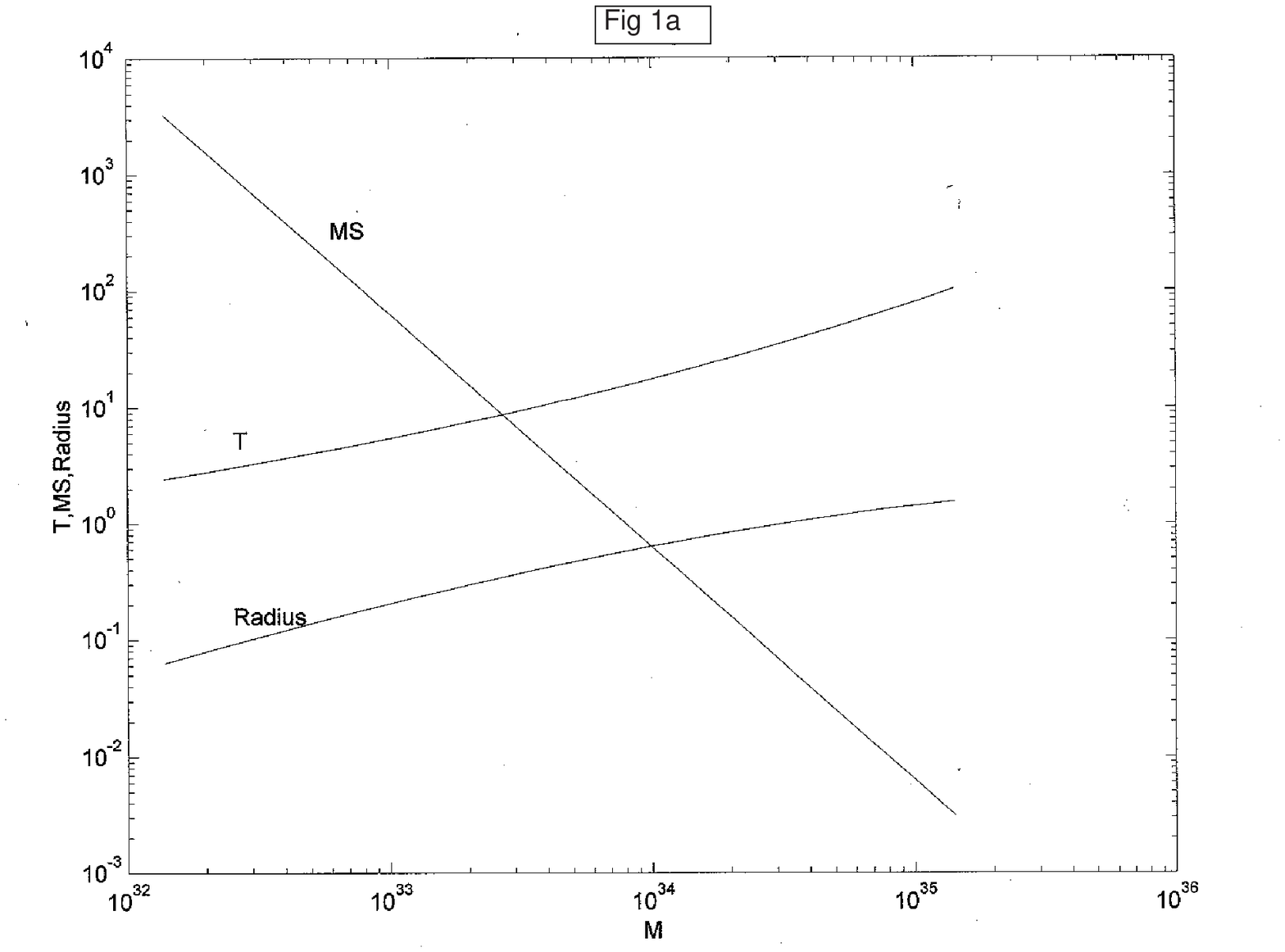}
\end{center}
\end{figure}

\begin{figure}[htb] \begin{center}
\epsfxsize=9.5cm
\epsfysize=9.5cm
\mbox{\hskip -1.0in}\epsfbox{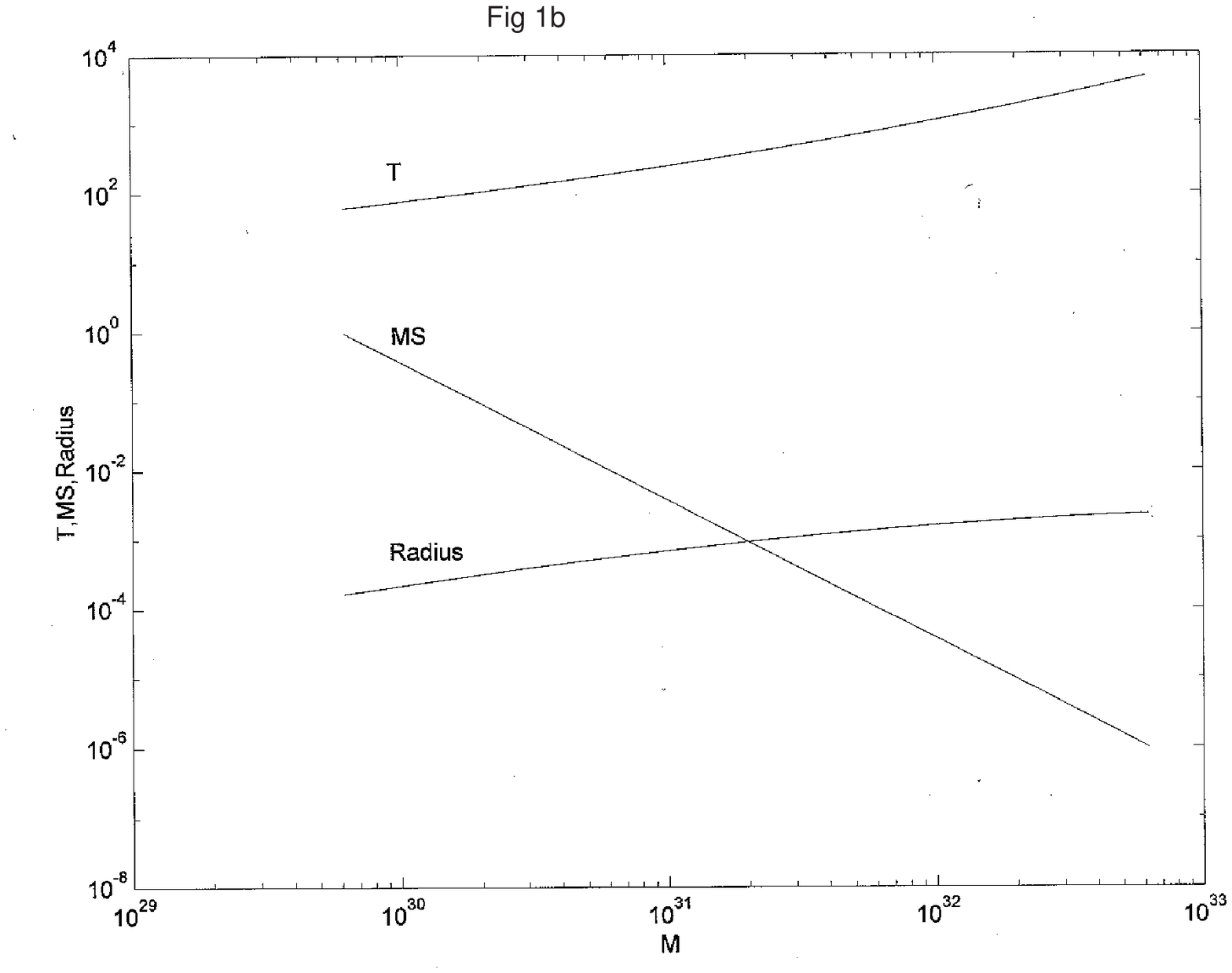}
\end{center}
\end{figure}

\begin{figure}[htb] \begin{center}
\epsfxsize=9.5cm
\epsfysize=9.5cm
\mbox{\hskip -1.0in}\epsfbox{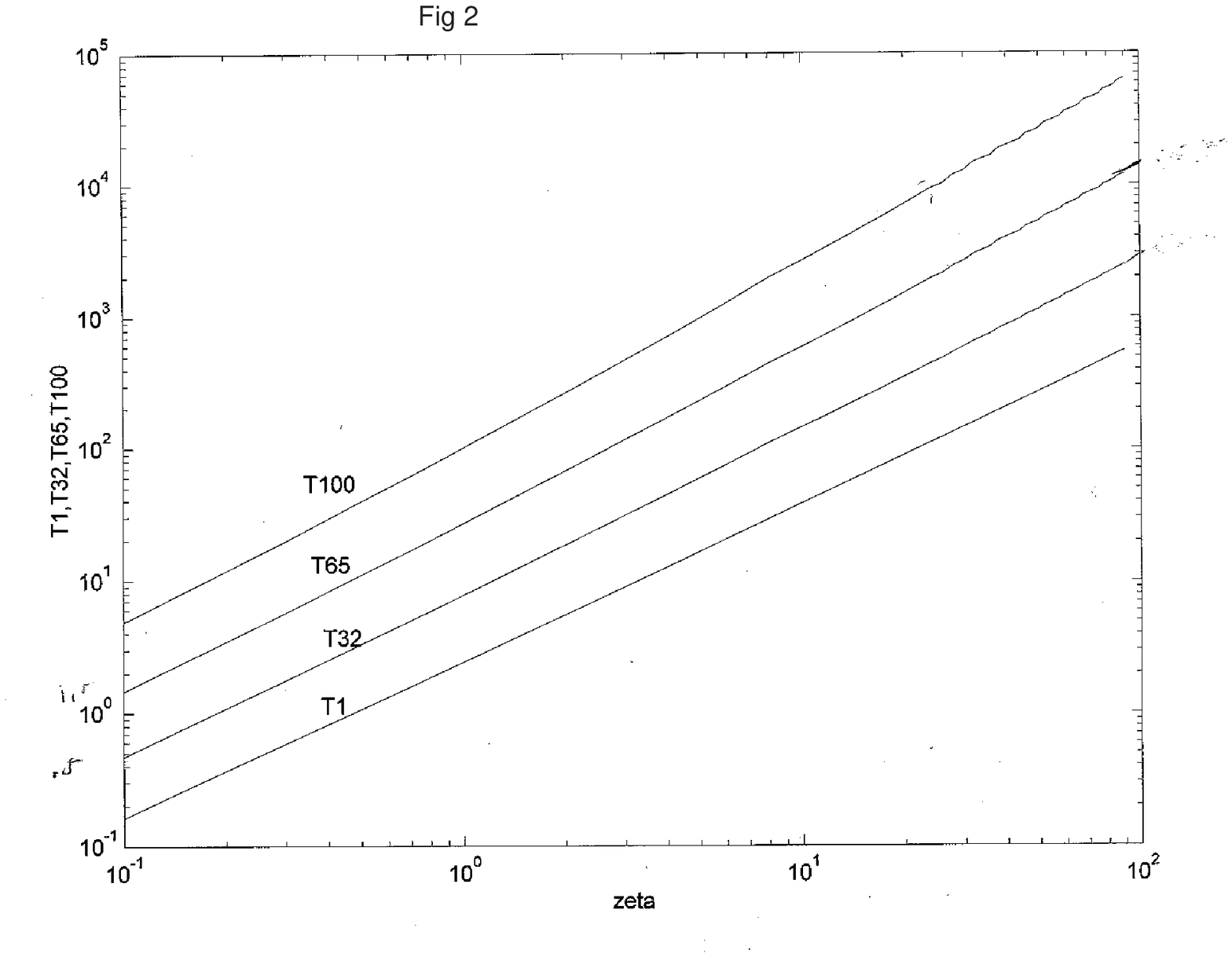}
\end{center}
\end{figure}

\end{document}